\documentclass[10pt,a4paper]{article}
 \pdfoutput=1
\usepackage{amsmath,fourier,amssymb,amsthm,graphicx,epstopdf,
color,cases,upgreek}
\usepackage{chngcntr}
\counterwithout{figure}{section}
\usepackage[T1]{fontenc}
\allowdisplaybreaks
\numberwithin{equation}{section}

   \newtheorem{innercustomthm}{Theorem}
   \newenvironment{customthm}[1]
     {\renewcommand\theinnercustomthm{#1}\innercustomthm}
     {\endinnercustomthm}

 \theoremstyle{plain}
 \newtheorem {hypo}{\bf\hspace{-\parindent}Hypothesis}[section]

 \newtheorem {prop}[hypo]{Proposition}

 \newtheorem {defin}[hypo]{Definition}
 
 \newtheorem {conj}[hypo]{Conjecture} 
 
 \theoremstyle{remark}
 \newtheorem {rmk}[hypo]{Remark}
  
 \newcommand{\pf}{\begin{bpf}}

 \newcommand{\pfms}{\begin{bpfms}}
 \newcommand{\epf}{\end{bpf}\hfill$\square$\vspace{0.1cm}}
 \newcommand{\epfms}{\end{bpfms}\hfill$\square$\\ }
 \newcommand\ben{\begin{equation*}}
 \newcommand\ebn{\end{equation*}}
 \newcommand\beq{\begin{equation}}
 \newcommand\eeq{\end{equation}}
 
  \newcommand\lb{\left(}
  \newcommand\rb{\right)}

   \newcommand\Cb{\mathbb{C}}

   \newcommand\sC{\mathsf{C}}
   \newcommand\sT{\mathsf{T}}

\def\bT{{\bf T}}

\def\bB{{\bf B}}
\def\bN{{\bf N}}
\def\bG{{\bf G}}

\def\bgam{\boldsymbol{\gamma}}
\def\ba{{\bf a}}
\def\bb{{\bf b}}
\def\bc{{\bf c}}
\def\Rb{\mathbb{R}}

\begin{document}
\LARGE
\noindent
\textbf{On self-similar solutions of the vortex filament equation}
\normalsize
 \vspace{1cm}\\
 \noindent\textit{
 O. Gamayun$\,^{a,}$\footnote{o.gamayun@uva.nl}, 
 O. Lisovyy$\,^{b,}$\footnote{lisovyi@idpoisson.fr}}
 \vspace{0.2cm}\\
$^a$ Institute for Theoretical Physics, Universiteit van Amsterdam
Science Park 904, Postbus 94485, \\1098 XH Amsterdam, The Netherlands
    \vspace{0.1cm}\\
 $^b$ Institut Denis-Poisson CNRS/UMR 7013,  Universit\'e de Tours, Parc de Grandmont, 37200 Tours, France
 \noindent 
 
 \begin{abstract}
 \noindent We study self-similar solutions of the binormal curvature flow which governs the evolution of vortex filaments and is equivalent to the Landau-Lifshitz equation. The corresponding dynamics is described by the real solutions of $\sigma$-Painlev\'{e} IV equation with two real parameters. Connection formulae for Painlev\'{e} IV transcendents allow for a complete characterization of the asymptotic properties of the curvature and torsion of the filament. We also provide compact hypergeometric expressions for self-similar solutions corresponding to corner initial conditions.
 \end{abstract}
  
 \section{Introduction and statement of results}
 The present note is concerned with the study of self-similar solutions
 of the geometric flow
 \beq\label{LIA}
 \bgam_t=\bgam_x\wedge\bgam_{xx},
 \eeq
 where $\bgam$ is a curve in $\Rb^3$ parameterized by its arc-length  $x$. The equation \eqref{LIA} was first considered more than a century ago by Da Rios \cite{DaRios,Ricca} and rederived in the middle 60s \cite{AH,Bet} as a simplified model for dynamics of a vortex filament in an inviscid incompressible fluid. In this context, it is often referred to as the localized induction approximation (LIA). 
 
 The Frenet orthonormal frame for $\bgam$ is formed by the tangent, normal and binormal vectors 
 $\bT=\bgam_x$, $\bN=\bT_x/\|\bT_x\|$ and $\bB=\bT\wedge\bN$.
  Denoting by $c$ and $\tau$ the curvature and torsion of the curve, the LIA can be equivalently rewritten as the binormal curvature flow, $\bgam_t=c\bB$. Yet another form is obtained by differentiating \eqref{LIA} with respect to $x$ and rewriting the result in terms of the tangent vector $\bT$,
  \beq
  \bT_t=\bT\wedge\bT_{xx},\qquad \|\bT\|=1.
  \eeq
  This is the Landau-Lifshitz (LL) equation for a one-dimensional continuous Heisenberg spin chain. The arc-length parameter $x$ plays therein the role of spatial coordinate and $\bT\lb x,t\rb$ is the spin field, which may be represented by a point on the  unit sphere $\mathbb S^2$.
  
  The self-similar solutions of \eqref{LIA} discussed in this work depend on  a constant vector $\ba\in\Rb^3$. Let $\operatorname{ad}_\ba$ be the linear transformation defined by
    $\operatorname{ad}_\ba{\bf v}={\bf a}\wedge{\bf v}$ for any 
    ${\bf v}\in\mathbb R^3$ and consider the ansatz
    \beq\label{SSt}
    \bgam\lb x,t\rb=\sqrt{t}\cdot t^{\frac12\operatorname{ad}_{\bf a}}\;\bG\lb s\rb, \qquad  
    s=\tfrac x{\sqrt t}.
    \eeq
  Hereafter it will be assumed that $t>0$ so that $s\in\Rb$. Note  that $\bT\lb x,t\rb= t^{\frac12\operatorname{ad}_\ba}\;\bG'\lb s\rb$. The above substitution  transforms the LIA equation
  \eqref{LIA} into a system of coupled nonlinear ODEs
  \beq\label{SSs}
  \ba\wedge\bG+\bG-s\bG'=2\bG'\wedge\bG'',\qquad \|\bG'\|=1,
  \eeq
  where the latter constraint follows from $\|\bT\|=1$ and the antisymmetry of $\operatorname{ad}_\ba$ (in other words, from the fact that $t^{\frac12\operatorname{ad}_{\bf a}}$ is a rotation). 
  It is worthwhile to note already at this point that \eqref{SSs} admits a nontrivial integral of motion. Indeed, computing the scalar and exterior product of this equation with $\bG'$, we find that ${\lb\ba\wedge\bG+\bG\rb\cdot \bG'=s}$ and $\bG''=\frac12\lb\ba\wedge\bG+\bG\rb\wedge \bG'$. It then becomes straightforward to check that the quantity
   \beq\label{epsI}
   \varepsilon=\tfrac14\left[\lb a^2+1\rb \|\bG\|^2-\lb \ba\cdot \bG\rb^2+4\lb\ba\cdot\bG'\rb-s^2\right],
   \eeq 
   where $a=\|\ba\|$, is conserved. In particular, for $a=0$ it is given by $\varepsilon=\frac14\lb\|\bG\|^2-s^2\rb=\|\bG''\|^2\ge0$ and coincides with the energy density of the LL ferromagnet.

\begin{figure}
      \centering
      \includegraphics[height=7cm]{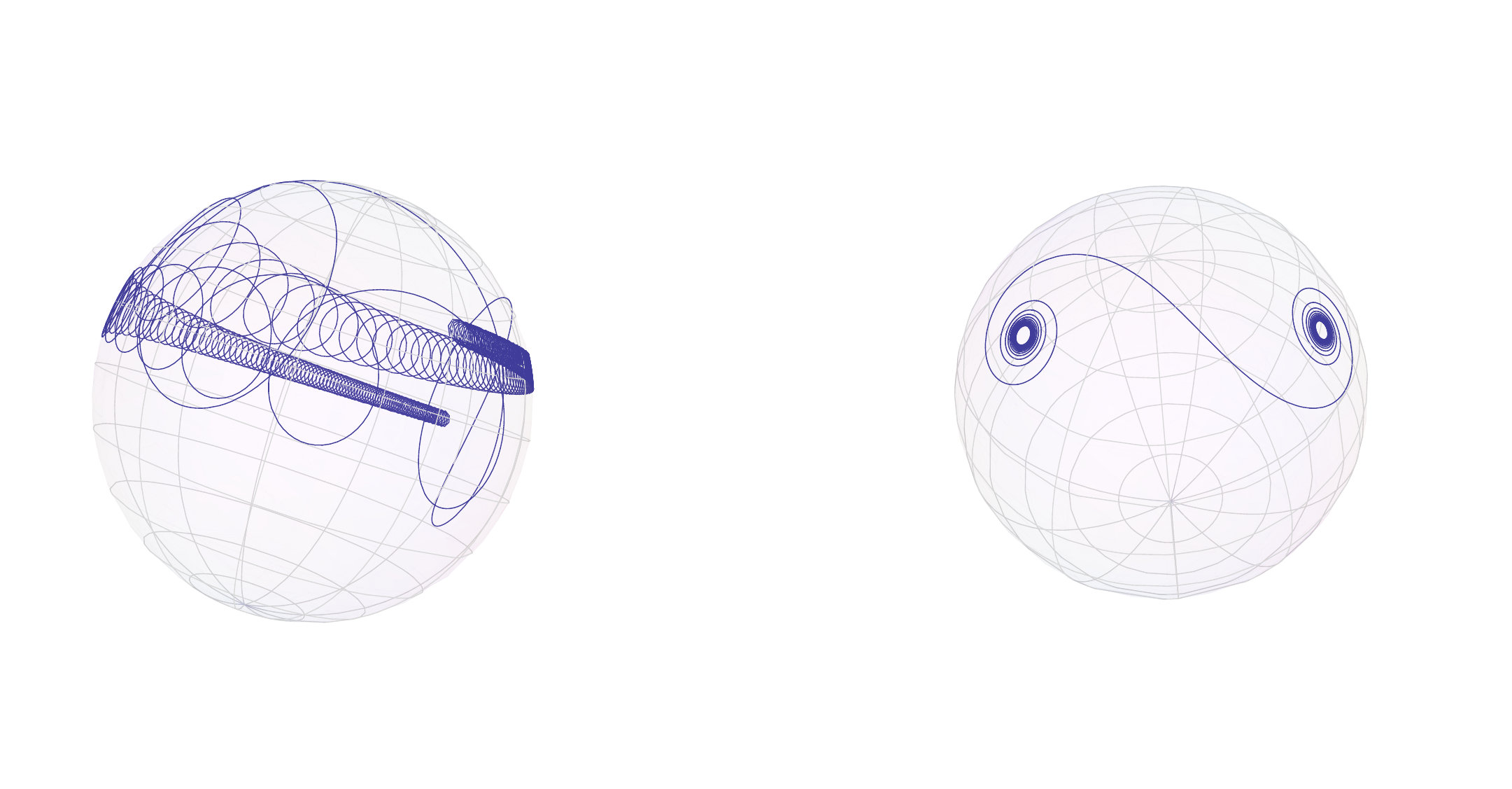}\vspace{-0.5cm}
    \caption{The path of the endpoint of $\bG'\lb s\rb$ on $\mathbb S^2$ for $a>0$ (left) and $a=0$ (right). \label{Fig1}}  
      \end{figure}

  The qualitative behavior of the self-similar solutions \eqref{SSt} is drastically different for $a>0$ and $a=0$, which is perhaps best seen in the asymptotics of $\bG'\lb s\rb$ as $s\to\pm\infty$. The effect of the distinguished direction $\ba$ is a logarithmically slowing precession of $\bG'\lb s\rb$ along two circles $\mathcal C_\pm$ orthogonal to $\ba$ (left part of Fig.~\ref{Fig1}). For $a=0$, the trajectory asymptotically approaches fixed points $\bT_\pm=\bG'\lb\pm\infty\rb$ along two spirals, as shown in the right part of Fig.~\ref{Fig1}.

  Self-similar solutions with $a=0$ were first considered in 
  \cite{LRT} in the context of the LL equation. Their asymptotics has been completely characterized by Guti\'errez, Rivas and Vega in \cite{GRV}. In this case, the curvature and torsion are given by particularly simple expressions, 
  \beq
  c\lb x,t\rb=\sqrt{\frac{\varepsilon}{t}},\qquad \tau\lb x,t\rb=\frac{x}{2t}.
  \eeq
  In fact, in  \cite{GRV} it is shown that equation \eqref{SSt} with $a=0$ can be reduced to linear ODEs whose solutions admit explicit integral representations. These latter allow not only to describe the asymptotics of $\bG\lb s\rb$ but also to extract explicit connection formulas between the asymptotic parameters of two spirals at $s\to\pm\infty$. The relevant solutions $\bG\lb s\rb$ are completely determined (up to rotations) by the value of the parameter $\varepsilon$. The stability of self-similar $a=0$ solutions has been extensively studied, see e.g. \cite{BV} and references therein.

 An analog of the ansatz \eqref{SSt} with $a>0$ was considered in \cite{Lip1,Lip2} for a dissipative deformation of the binormal flow. The systematic study of self-similar solutions \eqref{SSs} with $a>0$ was initiated in \cite{GV}; their stability properties were investigated in \cite{GV2}. As we will see shortly, for fixed $a$ and $\varepsilon$ and up to rotations around $\ba$, there is a 2-parameter family of solutions for $c$, $\tau$ and $\bG$. While the general form of the asymptotics of $\bG\lb s\rb$ as $s\to\pm\infty$ has been established already in \cite{GV}, the corresponding connection problem has not yet been solved; the challenge comes from the lack of explicit representations for solutions. Obtaining connection formulas between the parameters of asymptotic precessional motion at $s\to\pm\infty$  constitutes the main goal of this work.

 We are going to relate the self-similar dynamics of $\bgam$ for $a\ne 0$ to the fourth Painlev\'{e} (PIV) equation. Painlev\'{e} equations  describe monodromy preserving deformations of model systems of linear ODEs with rational coefficients \cite{JMU,JMII}. We refer the reader to \cite{Clarkson,FIKN} for detailed account of Painlev\'{e} theory, its earlier history and applications.

 \begin{customthm}{A} \label{ThA} The ``potential'' $\sigma\lb s\rb=\ba\cdot \bG\lb s\rb$ satisfies the $\sigma$-form of Painlev\'{e} IV equation:
 \beq\label{sP4}
  \lb \sigma''\rb^2+ \tfrac14\lb s\sigma'-\sigma\rb^2=\lb\sigma'-a\rb
  \lb\sigma'+a\rb\lb \sigma'-\varepsilon\rb.
 \eeq 
 The curvature and torsion of $\bgam$ are given by 
 \beq\label{C2T1}
 c\lb x,t\rb=\frac {1}{\sqrt t}\sC\lb \frac x{\sqrt t}\rb,\qquad 
 \tau\lb x,t\rb=\frac {1}{\sqrt t}\sT\lb \frac x{\sqrt t}\rb,
 \eeq
 where the functions $\sC\lb s\rb$, $\sT\lb s\rb$ are expressed in terms of $\sigma\lb s\rb$ as
 \beq\label{C2T2}
 \sC^2=\varepsilon-\sigma',\qquad 4\sC^2\lb\sT-\frac s2\rb=s\sigma'-\sigma.
 \eeq
 
 \end{customthm}
 
 Though this connection to PIV does not seem to be known to the LIA community, it can hardly be qualified as unexpected. It is well-known that the Hasimoto transformation \cite{Hasimoto}
   \beq\label{hasimoto}
   \psi\lb x,t\rb=c\lb x,t\rb \exp{ i\int^x\tau\lb x',t\rb dx'}
   \eeq
   relates the LIA flow to the nonlinear Schrodinger (NLS) equation, and that, on the other hand,  NLS admits a scaling reduction to generic PIV with two free parameters \cite{Can}. A special case of this reduction leading to a one-parameter PIV had been found earlier in \cite{BP}. The 2-parameter PIV was also obtained by Quispel and Capel \cite{QC} as a scaling reduction of the Landau-Lifshitz equation, and the first statement of Theorem A may be viewed as a reformulation of their results in invariant form. All of this is of course consistent with the general lore that the ODEs obtained as exact reductions of integrable PDEs should belong to Painlev\'{e} class.
 
  The general solution of PIV is, at least in some sense, transcendental, and the reader might wonder whether the identification of the equation has any benefit. One may recall that for special parameter values and initial conditions PIV admits  explicit rational and special function solutions \cite{BCHM}. We will see, however, that these solutions are irrelevant for LIA. Secondly, and more importantly, the connection problem has been effectively solved for generic PIV transcendent by Kitaev \cite{Kitaev}  and Kapaev \cite{Kapaev} by expressing the asymptotic parameters in terms of monodromy data of the associated linear system. This leads to
 
  \begin{customthm}{B} \label{theob} The asymptotic behavior of $\mathsf C\lb s\rb$ and $\mathsf T\lb s\rb$ as $s\to\pm\infty$ satisfies
  \begin{subequations}\label{asyCT}
  \begin{align}
   &\mathsf C^2\lb s\rb\simeq\frac{2\lb \varepsilon -3\omega_\pm\rb}{3}\mp\frac{2R\lb\omega_\pm\rb\cos\lb\frac{s^2}{4}-6\omega_\pm\ln\tfrac{|s|}{\sqrt2}+\delta_\pm\rb}{9s}+O\lb s^{-2}\rb,\\
   &\lb\varepsilon-3\omega_\pm\rb\lb \mathsf T\lb s\rb-\frac s2\rb\simeq \pm
   \frac{R\lb\omega_\pm\rb}{12}\cos\lb\tfrac{s^2}{4}-6\omega_\pm\ln\tfrac{|s|}{\sqrt2}+\delta_\pm\rb+O\lb s^{-1}\rb,
  \end{align}
  \end{subequations}
  where $R\lb\omega\rb=\sqrt{6\lb\varepsilon-3\omega\rb \lb 9a^2-\lb\varepsilon+6\omega\rb^2\rb}$ and $\omega_\pm,\delta_\pm$ are real parameters satisfying $-\frac a2-\frac{\varepsilon}{6}\leq\omega_\pm\leq\min\left\{\frac{\varepsilon}{3},\frac a2-\frac{\varepsilon}{6}\right\}$. The connection formulas relating the pairs $\lb\omega_+,\delta_+\rb$ and $\lb\omega_-,\delta_-\rb$ are given by
  \begin{subequations}
  \label{connfI}
  \begin{align}
  & e^{-2\pi\omega_\mp}=2e^{4\pi\omega_\pm}\lb e^{-\Im \rho_\pm}\cos\Re\rho_\pm-1\rb+e^{2\pi\omega_\pm}\lb 2e^{-\frac{\pi\varepsilon}{3}}\cosh\pi a+e^{\frac{2\pi\varepsilon}{3}}\rb,\\
  & e^{2\pi\omega_+}\lb 1-e^{i\rho_+}\rb+e^{2\pi\omega_-}\lb 1-e^{i\rho_-}\rb=2e^{-\frac{\pi\varepsilon}{3}}\cosh\pi a+e^{\frac{2\pi\varepsilon}{3}} -e^{-2\pi\lb\omega_++\omega_-\rb},
  \end{align}
  \end{subequations}
  where
  \begin{subequations}
  \label{connfII}
    \begin{gather}
    \Re\rho_\pm=\delta_\pm-\arg\Gamma\lb1+\tfrac{i\lb\varepsilon+6\omega_\pm-3a\rb}{6}\rb-\arg\Gamma\lb1+\tfrac{i\lb\varepsilon+6\omega_\pm+3a\rb}{6}\rb-\arg\Gamma\lb1+\tfrac{i\lb3\omega_\pm-\varepsilon\rb}{3}\rb+\tfrac{3\pi}{4},\\    
    e^{-2\Im\rho_\pm}=4e^{-3\pi\omega_\pm}\sinh\tfrac{\pi\lb\varepsilon-3\omega_\pm\rb}{3}\lb\cosh\pi a-\cosh\tfrac{\pi\lb\varepsilon+6\omega_\pm\rb}{3}\rb.
    \end{gather}
 \end{subequations}   
  \end{customthm}  
  As already mentioned, self-similar LIA solutions with $a=0$ were constructed in terms of contour integrals. The corresponding expressions are not easy to identify in \cite{GRV}; it actually took us a while to realize that Section~3 therein effectively solves the problem. We therefore decided to include to the present note a simple derivation of the $a=0$ solution as well as its explicit compact expression in terms of the confluent hypergeometric and parabolic cylinder functions, which has been overlooked in \cite{GRV}.
  
     Consider the self-similar flow \eqref{SSs} with $a=0$:
     \beq\label{ae0eq}
    \bG-s\bG'=2\bG'\wedge\bG'',\qquad \|\bG'\|=1.   
     \eeq 
    Fix an orthonormal frame $\lb\mathbf{e}_1,\mathbf{e}_2,\mathbf{e}_3\rb$. Recalling that $\varepsilon=\|\bG''\|^2$ and using rotational invariance, it may be assumed without loss of generality that the initial conditions for \eqref{ae0eq} at $s=0$ are given by
    \beq\label{ae0ic}
    \bG'\lb 0\rb=\lb 1,0,0\rb,\qquad \bG''\lb 0\rb=\lb 0,\sqrt{\varepsilon},0\rb.
    \eeq
    To solve the corresponding Cauchy problem, it suffices to find the expression of  $\bG'\lb s\rb$; it turns out to be simpler than that of $\bG\lb s\rb$, which itself can be easily recovered from  \eqref{ae0eq}. Observe that under the above choice of initial conditions $G'_1\lb s\rb$ is an even function of $s$ while $G'_{2,3}\lb s\rb$ are odd.
 
  \begin{customthm}{C} \label{theoc} 
  The solution of \eqref{ae0eq} subject to the initial conditions \eqref{ae0ic} is given by
    \begin{subequations}\label{1F1eqs}
  \begin{gather}
  G_1'\lb s\rb=1-\tfrac{\varepsilon s^2}{2}{\,}_1F_1\lb \tfrac12+\tfrac{i\varepsilon}{4},\tfrac32,\tfrac{is^2}{4}\rb{\,}
  _1F_1\lb \tfrac12-\tfrac{i\varepsilon}{4},\tfrac32,-\tfrac{is^2}{4}\rb,\\
  G_2'\lb s\rb\pm iG_3'\lb s\rb=\sqrt\varepsilon\,s {\,}
  _1F_1\lb \tfrac12\pm\tfrac{i\varepsilon}{4},\tfrac32,\pm\tfrac{is^2}{4}\rb{\,}
    _1F_1\lb \mp\tfrac{i\varepsilon}{4},\tfrac12,\mp\tfrac{is^2}{4}\rb,
  \end{gather}   
    \end{subequations}
  where $_1F_1\lb \alpha,\gamma,z\rb$ denotes the confluent hypergeometric function.   
  \end{customthm}

 The paper is organized as follows. In the next section, we show how the self-similar LIA flow \eqref{SSs} reduces to Painlev\'{e} IV. Section~\ref{sec_conn} describes the monodromy data relevant to self-similar LIA solutions and characterizes the geometric properties of $\bG\lb s\rb$ as $s\to\pm\infty$ in terms of these data. The special case of symmetric solutions, for which the asymptotics at $s\to\pm\infty$ can be related to the Cauchy data at $s=0$, is discussed in Subsection~\ref{subsecodd}. The derivation of special function representations of $a=0$ solution is outlined in Section~\ref{sec_ae0}.

 \section{Self-similar dynamics and Painlev\'{e} IV}
 The goal of this section is to establish the relation between the self-similar solutions \eqref{SSt} of LIA and Painlev\'{e} IV functions outlined in Theorem~\ref{ThA}. Recall the definitions $\sigma\lb s\rb=\ba\cdot\bG\lb s\rb$ and $a=\|\ba\|$. It will be assumed that $a\ne 0$ unless  otherwise specified. 
 
 Differentiating \eqref{SSs} once with respect to $s$, we find the equation
 \beq\label{SSt2}
 \ba\wedge\bG'-s\bG''=2\bG'\wedge\bG'''.
 \eeq
 Computing the scalar product of this equation with $\bG'''$ and using that $\lb\ba\wedge\bG'\rb\cdot\bG'''=\ba\cdot\lb \bG'\wedge\bG'''\rb=-\tfrac12 s\sigma''$, it follows that 
 \beq\label{epsI2}
 \varepsilon=\|\bG''\|^2+\sigma'
 \eeq
 is an integral of motion for \eqref{SSs}. This quantity coincides with $-\alpha$ in \cite{GV}. In order to check that it can be equivalently written as \eqref{epsI}, it suffices to use the identities $\bG''=\frac12\lb\ba\wedge\bG+\bG\rb\wedge \bG'$ and $\|\bG'\|=1$.  On the other hand, the scalar product of \eqref{SSs} with $\ba$ yields
 \beq\label{mixed}
 \ba\cdot\lb \bG'\wedge\bG''\rb=\frac12\lb \sigma-s\sigma'\rb.
 \eeq
 
 Any triple of vectors $\ba,\bb,\bc\in\Rb^3$ satisfies the identity
   \beq\label{determ}
   \lb\ba\cdot\lb\bb\wedge\bc\rb\rb^2=\operatorname{det}
   \lb\begin{array}{ccc}
   \ba\cdot\ba & \ba\cdot\bb & \ba\cdot\bc \\
   \bb\cdot\ba & \bb\cdot\bb & \bb\cdot\bc \\ 
   \bc\cdot\ba & \bc\cdot\bb & \bc\cdot\bc   
   \end{array}\rb.
   \eeq
   Now, choosing
   $\bb=\bG'$, $\bc=\bG''$, we may express all entries of the symmetric matrix on the right in terms of $\sigma$ and its derivatives. Indeed, 
   \beq
   \|\ba\|^2=a^2,\quad\|\bG'\|^2=1,\quad \|\bG''\|^2=\varepsilon-\sigma', \quad \ba\cdot\bG'=\sigma',\quad \ba\cdot\bG''=\sigma'',\quad \bG'\cdot\bG''=0.
   \eeq
   Similarly, equation \eqref{mixed} provides an expression for the left-hand side of the identity \eqref{determ}. This yields a nonlinear 2nd order ODE for $\sigma$,
  \beq
  \tfrac14\lb \sigma-s\sigma'\rb^2=\operatorname{det}
  \lb\begin{array}{ccc}
  a^2 & \sigma' & \sigma'' \\
  \sigma' & 1 & 0 \\ \sigma'' & 0 & \varepsilon-\sigma'
  \end{array}
  \rb.
  \eeq
  which is clearly equivalent to the Painlev\'{e} IV equation \eqref{sP4} of Theorem~\ref{ThA}.

  The expressions \eqref{C2T1}--\eqref{C2T2} for the curvature and torsion can be deduced from the Frenet-Serret formulas $\bT_x=c\bN$, $\bN_x=-c\bT+\tau\bB$. Indeed, these formulas imply that 
  \beq
  c=\|\bT_x\|=\frac{\|\bG''\|}{\sqrt t},\qquad
  c^2\tau=\bT\cdot\lb\bT_x\wedge\bT_{xx}\rb=
  -\frac{\bG''\cdot\lb\bG'\wedge\bG'''\rb}{t\sqrt t}.
  \eeq 
  The first of expressions \eqref{C2T2} now immediately follows from \eqref{epsI2}. In order to prove the second one, it suffices to compute the scalar product of \eqref{SSt2} with $\bG''$ and use again \eqref{epsI2} and \eqref{mixed} to compute $\|\bG''\|^2$ and $\ba\cdot\lb \bG'\wedge\bG''\rb$. This ends the proof of Theorem~\ref{ThA}.
  
  \begin{rmk}
  It is far from trivial to characterize the PIV initial conditions relevant for the LIA flow. Here are some constraints. From $\|\bG'\|=1$ it follows that $|\sigma'\lb s\rb|\leq a$ for all $s\in\Rb$. This in turn implies that $\varepsilon\ge -a$. Similarly,  $\lb\ba\wedge \bG+\bG\rb\cdot\bG'=s$ implies that $\|\ba\wedge \bG+\bG\|\leq |s|$. The conserved quantity $\varepsilon$ is related to the length of $\|\bG\|$, which may be used to show that $a^{-2}\sigma^2-s^2+4\sigma'-4\varepsilon\ge0$. It is a classical fact that any local solution of PIV meromorphically continues to the whole complex $s$-plane. The LIA-PIV initial conditions are such that the poles do not appear on the real axis.
  \end{rmk}

  In order to express the projection of $\bG$ onto the plane orthogonal to $\ba$  in terms of $\sigma$, it is convenient to fix an orthonormal frame in which $\ba=\lb 0,0,a\rb$. Then, parameterizing $\bG'$ by spherical angles so that  $\bG'=\lb \cos\phi\sin\theta,\sin\phi\sin\theta,\cos\theta\rb$, one finds from \eqref{SSt2} the following equations of motion:
  \begin{subequations}
  \begin{align}\label{sphe1a}
  &2\theta''=\sin\theta\left[ 2\cos\theta\, \lb\phi'\rb^2-s\phi'+a\right],\\
  \label{sphe1b}
  &2\sin\theta\,\phi''=\lb s-4\cos\theta\,\phi'\rb\theta'.
  \end{align}
  \end{subequations}
  The conserved quantity $\varepsilon$ is expressed in terms of $\theta$, $\theta'$ and $\phi'$ as
  \beq\label{sphe2}
  \varepsilon=\lb\theta'\rb^2+\sin^2\theta\,\lb\phi'\rb^2+a\cos\theta.
  \eeq
  Recall that $\sigma'=a\cos\theta$. An alternative way to derive the $\sigma$-form \eqref{sP4} of the PIV equation is to eliminate $\phi'$ from \eqref{sphe1a} using \eqref{sphe2} . A compact expression for $\phi$ is most easily deduced from  \eqref{mixed}: using that $\ba\cdot\lb\bG'\wedge\bG''\rb=a\sin^2\theta \,\phi'$ we find that
  \beq
  \phi\lb s\rb=\frac{a}{2}\int_{s_0}^s\frac{s\sigma'-\sigma}{\lb\sigma'\rb^2- a^2}ds.
  \eeq
  The value of $s_0$ can be fixed arbitrarily; its different choices correspond to rotations around $\ba$.
  
  Before we move on to the asymptotics, let us recall the relation between the $\sigma$-PIV equation \eqref{sP4} and the standard form of PIV. Define the functions $q\lb s\rb$, $p\lb s\rb$ by
  \begin{subequations}\label{qpdef}
  \begin{align}
  \label{qdef}
  e^{-i\pi/4}q\lb\tfrac12 e^{-i\pi/4}s\rb=&\,-\frac{\sigma''+\frac i2\lb s\sigma'-\sigma\rb}{a-\sigma'},\\
  \label{pdef}
   e^{-i\pi/4}p\lb\tfrac12 e^{-i\pi/4}s\rb=&\,-\frac{\sigma''-\frac i2\lb s\sigma'-\sigma\rb}{a+\sigma'}.
  \end{align}
  \end{subequations}
  It is straightforward to check that $q\lb s\rb$ satisfies the conventional PIV equation
  \beq\label{cP4}
   q''=\frac{\lb q'\rb^2}{2q}+\frac32 q^3+4sq^2+2\lb s^2-\alpha\rb q+\frac{\beta}{q},
  \eeq
  with $\alpha=1-\frac{i\lb\varepsilon-3a\rb}{2}$, $\beta=\frac{\lb a+\varepsilon\rb^2}{2}$, whereas $p\lb s\rb$ satisfies the same equation with $\alpha=-1-\frac{i\lb\varepsilon+3a\rb}{2}$, $\beta=\frac{\lb a-\varepsilon\rb^2}{2}$.  Conversely, $\sigma$ can be expressed in terms of $q$; up to suitable rescalings, it coincides with the non-autonomous Hamiltonian for \eqref{cP4}. 
  
  \begin{rmk}
   It is known (see, for example, \cite[Theorem 6.1]{Clarkson}) that Painlev\'{e} IV equation \eqref{cP4} admits special function solutions if and only if
   $\beta=-2n^2$ or $\beta=-2\lb 2n+1\pm\alpha\rb^2$ with $n\in\mathbb Z$. Given that $a,\varepsilon\in\Rb$, in our case these constraints imply that $\varepsilon=-a$ or $\varepsilon=a$ and simultaneously fix the value of $n$. From \eqref{epsI} it follows that for $\varepsilon=-a$ one has $\bG''=0$, $\bG'=-\ba/a$, which leads to a trivial LIA solution $\bG\lb s\rb=-s\ba/a$. Let us examine the remaining possibility $\varepsilon=a$. In this case, the equation for $q\lb s\rb$ defined by \eqref{qdef} becomes
   \beq
   q''=\frac{\lb q'\rb^2}{2q}+\frac32 q^3+4sq^2+2\lb s^2-1-ia\rb q+ \frac{2a^2}{q}.
   \eeq 
   This admits a one-parameter family of solutions
   \beq
   q\lb s\rb=-s+\frac{d}{ds}\ln\lb C_+D_{ia}\lb is\sqrt2\rb+C_-D_{ia}\lb -is\sqrt2\rb\rb,
   \eeq
   where $D_\alpha\lb z\rb$ denotes the parabolic cylinder function. The initial conditions are parameterized here by the ratio~$C_+/C_-$. While at first sight such a solution may look non-trivial, the corresponding $\sigma$-function is given by $\sigma\lb s\rb=a s$ and  describes the same straight line $\bG\lb s\rb=s\ba/a$.
  \end{rmk}

  \section{Asymptotics and connection problem\label{sec_conn}}

   \subsection{Asymptotics}
     We have seen that, in the context of LIA, the argument of conventional (but inconvenient!) Painlev\'{e} function $q\lb s\rb$ lives on the lines 
      $s\in e^{\pm i\pi/4}\Rb$. The argument of $\sigma\lb s\rb$ belongs to the real (for $t>0$) or imaginary ($t<0$) axis. The asymptotics of PIV functions along the corresponding 4 directions at $\infty$ has been found by Kitaev in \cite{Kitaev}; his results have been completed by Kapaev \cite{Kapaev} who described the large $s$ asymptotics for arbitrary complex argument.
   
   In order to explain the relevant results of \cite{Kapaev,Kitaev}, it is useful to mention a conjecture which simultaneously generalizes and reformulates them in a compact way.
   Let us introduce the Painlev\'{e} IV \textit{tau function} $\uptau\lb s\rb$ (not to be confused with the torsion $\tau\lb x,t\rb$ of $\bgam$) by
   \beq\label{deftau}
   \sigma\lb s\rb=-4\frac{d}{ds}\ln\uptau\lb s\rb.
   \eeq
   This function holomorphically continues to the entire complex $s$-plane. Its zeros correspond to simple poles of $\sigma\lb s\rb$.
   The conjecture is as follows.
   \begin{conj}\label{mainconj}
   The asymptotic expansions of the generic PIV tau function $\uptau\lb s\rb$ as $s\to\infty$ along the rays ${\arg s=\frac{i\pi \ell}{2}}$, $\ell=0,\ldots,3$ can be represented in the form
   \beq\label{taufourier}
   \uptau\lb s\rb\simeq \Upsilon_\ell\sum_{n\in\mathbb Z}e^{in\rho_\ell}\mathcal D\lb i\omega_\ell+n,s\rb,
   \eeq
   where $\mathcal D\lb i\omega,s\rb$ expands as
   \begin{align}
   &\mathcal D\lb i\omega,s\rb\simeq C\lb i\omega\rb s^{3\omega^2-\frac{\varepsilon^2+3a^2}{12}}e^{-\frac{ \varepsilon+6\omega }{24} s^2} \left[1+\sum_{k=1}^{\infty}\frac{ \mathcal D_k\lb i\omega\rb}{s^{2k}}\right],\\
   \label{strc}
   &C\lb i\omega\rb=2^{-\frac{3\omega^2}{2}}e^{\frac{3i\pi\omega^2}{4}}
   \lb2\pi\rb^{-\frac{3i\omega}{2}}
   G\lb 1+i\omega +\tfrac{i\varepsilon}{6}-\tfrac{ia}{2}\rb
     G\lb 1+i\omega +\tfrac{i\varepsilon}{6}+\tfrac{ia}{2}\rb
     G\lb 1+i\omega -\tfrac{i\varepsilon}{3}\rb,
   \end{align}
  and $G\lb z\rb$ is the Barnes $G$-function. 
   \end{conj}
   A proposal essentially equivalent to the above has been first put forward by Nagoya \cite{Nagoya1,Nagoya2}. It extends earlier similar results for Painlev\'{e} VI, V and III \cite{GIL12,GIL13,ILT}.  Fourier expansions of type \eqref{taufourier} have been found in \cite{BLMST} for tau functions of all Painlev\'{e} equations on different sets of canonical rays. They have the meaning of resurgent trans-series encoding the sums over all multi-instanton sectors, labeled by $n$ in \eqref{taufourier}, and fluctuations inside each sector \cite{Dunne}. The pairs of parameters $\lb\omega_\ell,\rho_\ell\rb$ represent PIV initial conditions.
      
  Let us make a few comments:
   \begin{itemize}
   \item The qualifier ``generic'' in Conjecture~\ref{mainconj} will be clarified below in terms of monodromy. For the moment, we just note that this includes all solutions with 2-parameter dependence on the initial conditions, such as those described in \cite[Theorem 5]{Kapaev}. Painlev\'{e} IV also admits 0- and 1-parameter families of degenerate solutions \cite{Kapaev,KapaevII} whose asymptotics may differ from \eqref{taufourier}--\eqref{strc}, but they are not relevant for LIA.
   \item
   It is not difficult to understand that the asymptotics of $\sigma\lb s\rb$ depends on the structure constants  only via the ratios
   $C\lb i\omega+m\rb/C\lb i\omega+n\rb$ with $m,n\in\mathbb Z$. Hence the only property of $G\lb z\rb$ relevant for us is the recurrence relation $G\lb z+1\rb=\Gamma\lb z\rb G\lb z\rb$ expressing such ratios in terms of gamma functions.
   \item The coefficients $\mathcal D_k\lb i\omega\rb$ are polynomial in $\omega$, $\varepsilon$, $a$. They may be straightforwardly computed using the differential equation \eqref{sP4}. In particular,
   \beq
   \mathcal D_1\lb i\omega\rb=-12\omega^3+\frac{\varepsilon^2+3a^2}{2}\omega+
   \frac{\varepsilon\lb\varepsilon^2-9a^2\rb}{36}.
   \eeq 
   \item The asymptotic series $\mathcal D\lb i\omega,s\rb$ has a representation-theoretic meaning: in \cite{Nagoya1,Nagoya2}, it was introduced as a specialization of an irregular conformal block of the Virasoro algebra  which involves a vertex operator intertwining two rank 2 Whittaker modules. A direct algebraic computation of $\mathcal D_k$ from the Virasoro commutation relations is possible but it becomes rather cumbersome as $k$ increases.
   \end{itemize}
   
 The periodic structure of \eqref{taufourier} allows one to assume that every $\omega=\omega_\ell$ belongs to the strip $\Im \omega\in(-\frac12,\frac12]$. Assuming the validity of Conjecture~\ref{mainconj}, the asymptotics of $q\lb s\rb$ on the rays $\arg s=-\tfrac{\pi}4+\tfrac{\pi \ell}{2}$ can be derived from that of $\uptau\lb s\rb$ using \eqref{deftau} and \eqref{qdef}. Its form is more involved than of $\uptau\lb s\rb$ and depends on the value of $\Im \omega$ (cf different cases in \cite{Kapaev,Kitaev}) because of the need to compute the logarithmic derivatives. For example, for $|\Im \omega|<\frac16$ one has
 \begin{subequations}\label{sigmaas}
 \begin{align}
 \label{sigmaas0}
 \sigma\lb s\rb=&\,\frac{\varepsilon+6\omega}{3} s+\frac{a^2-12\omega^2+\frac{\varepsilon^2}{3}}s-
 \frac{2i}{s^2}\lb As^{-6i\omega}e^{\frac{is^2}{4}}-
 Bs^{6i\omega}e^{-\frac{is^2}{4}}\rb-\frac{8\mathcal D_1\lb i\omega\rb}{s^3} 
 +O\lb|s|^{6|\Im\omega|-4}\rb,\\
 \label{sigmapas}
 \sigma'\lb s\rb=&\,\frac{\varepsilon+6\omega}{3}+\frac{ As^{-6i\omega}e^{\frac{is^2}{4}}+  Bs^{6i\omega}e^{-\frac{is^2}{4}}}{s}-\frac{a^2-12\omega^2+\frac{\varepsilon^2}{3}}{s^2}
 +O\lb|s|^{6|\Im\omega|-3}\rb,\\
 \sigma''\lb s\rb=&\,\frac i2\lb As^{-6i\omega}e^{\frac{is^2}{4}}-  Bs^{6i\omega}e^{-\frac{is^2}{4}}\rb+O\lb|s|^{6|\Im\omega|-2}\rb,
 \end{align}
 \end{subequations}
 where $A=e^{i\rho}C\lb i\omega+1\rb/C\lb i\omega\rb$ and $B=e^{- i\rho}C\lb i\omega-1\rb/C\lb i\omega\rb$.  These coefficients satisfy 
 \beq\label{cpcm}
   AB=\frac{2\lb\varepsilon-3\omega\rb \lb 9a^2-\lb\varepsilon+6\omega\rb^2\rb}{27}.
 \eeq
  If we now assume in addition that $0\leq\Im \omega<\frac16$, then it follows from the above estimates and the definition \eqref{qdef} of $q\lb s\rb$ that
 \beq\label{asq}
 \frac{e^{i\pi/4}}{q\lb\tfrac12 e^{-i\pi/4}s\rb}=\frac{\varepsilon+6\omega-3a}{3iA}\,s^{6i\omega}e^{-\frac{is^2}{4}}+O\lb|s|^{-1}\rb.
 \eeq  
  
  The asymptotics \eqref{asq} matches the results of \cite{Kapaev,Kitaev}; see, for instance,  the cases (ii) (for $s\to\pm i\infty$) and (iii) (for $s\to\pm \infty$) of Theorem~5 in \cite{Kapaev}. On the other hand, once the leading asymptotics is known, it is not difficult to calculate subleading corrections to it  to any desired order from the differential equation, and it is this computation which leads to Conjecture~\ref{mainconj}. Although complete asymptotic expansion for $\uptau\lb s\rb$ (in particular, the periodic structure of its coefficients) has not yet been proved, the proof of correctness of its truncation to any finite number of terms is a routine procedure. 
  
  We are interested in generic solutions of $\sigma$-PIV equation \eqref{sP4} which are real for $s\in\mathbb R$. Such solutions can exist only if $\Im \omega=0$ or $\frac12$ ($\operatorname{mod} \mathbb Z$). It turns out that the second option is inconsistent with the LIA asymptotics of $\bG\lb s\rb$ found in \cite[Theorem 1]{GV} ($\sigma\lb s\rb=aG_3\lb s\rb$ in their notation). One reason for this is that $\sigma\lb s\rb$ in this case develops an infinite number of poles on the real axis. We may therefore assume that $\Im \omega=0$. Also, since $\sigma'\lb s\rb\in[-a,a]$, from the asymptotics \eqref{sigmapas} it follows that $-\frac a2-\frac{\varepsilon}6\leq\omega\leq\frac a2-\frac{\varepsilon}6$. 
  
  Let us introduce the notation 
  \beq
  \omega_+=\omega_0,\qquad \omega_-=\omega_2,\qquad
  \rho_+=\rho_0,\qquad \rho_-=\rho_2+\pi\lb1-6i\omega_2\rb.
  \eeq
  The parameters $\omega_\pm$ are directly related to the positions of asymptotic circles $\mathcal C_{\pm}$ by $\sigma'\lb\pm\infty\rb=\ba \cdot \bG'\lb \pm\infty\rb=\frac{\varepsilon+6\omega_{\pm}}{3}$, as well as to limiting values $\mathsf C\lb\pm\infty\rb$ of the curvature scaling function by \eqref{C2T2}. Note that the equality $\varepsilon-\sigma'=\|\bG''\|^2$, which remains valid for all $s$, implies that $\varepsilon\ge3\omega_\pm$, and therefore the product \eqref{cpcm} is non-negative for $\omega=\omega_\pm$.
  
  Subleading corrections in \eqref{sigmaas} should also be real, therefore we have to require that $A=B^*$. This fixes the imaginary part of $\rho_\pm$:
  \beq
  e^{-2\Im\rho_\pm}=4e^{-3\pi\omega_\pm}\sinh\tfrac{\pi\lb\varepsilon-3\omega_\pm\rb}{3}\lb\cosh\pi a-\cosh\tfrac{\pi\lb\varepsilon+6\omega_\pm\rb}{3}\rb.
  \eeq
  The real part of $\rho_\pm$ remains unconstrained and appears in the phase shifts $\delta_\pm=\arg A_\pm$.
  This finishes the derivation of the asymptotics \eqref{asyCT} in Theorem~\ref{theob}. 
  
  Any of the pairs $\lb\omega_+,\delta_+\rb$, $\lb\omega_-,\delta_-\rb$ of asymptotic parameters at $s\to \pm\infty$ fixes the initial conditions for \eqref{sP4}. The connection formulas between the two pairs can be obtained using the monodromy data of the linear system associated to PIV, which we describe in the next subsection.
  
  \subsection{PIV monodromy data and connection formulas}
  
  Painlev\'e IV naturally arises in the description of isomonodromic deformations of rank $2$ linear systems with one regular singularity and one irregular singularity of Poincar\'e rank $2$ on the Riemann sphere $\mathbb C\mathbb P^1$. However, the papers \cite{Kapaev,Kitaev} use instead a system with one regular and one irregular point of rank $4$ which admits a particular $\mathbb Z_2$-symmetry. This latter system comes from the standard Lax pair for NLS equation and is explicitly given by
  \beq\label{lisys}
  \partial_z\Phi=A\lb z\rb\Phi,\qquad A\lb z\rb=\lb\begin{array}{cc}
    \frac{z^3}{2}+z\lb s+uv\rb+\theta_0 z^{-1} & 
    i\lb z^2u+2s u+u_1\rb\\
    i\lb z^2v+2s v-v_1\rb & -\frac{z^3}{2}-z\lb s+uv\rb-\theta_0 z^{-1}
    \end{array}\rb.
  \eeq
  The discrete symmetry in question is $A\lb -z\rb=-\sigma_3A\lb z\rb \sigma_3$, where $\sigma_3=\operatorname{diag}\left(1,-1\right)$. It implies that if $\Phi\lb z\rb$ is a solution of \eqref{lisys}, then so is $\sigma_3\Phi\lb -z\rb$. The quantities $s$, $\theta_0$, $u$, $v$, $u_1$, $v_1$ parameterize the space of systems with such symmetry and prescribed singularity structure.
 
 If the regular singularity is non-resonant (i.e. $\theta_0\notin\mathbb Z+\frac12$), then there exists a fundamental matrix solution $\Phi^{(0)}\lb z\rb$ characterized by diagonal monodromy at $z=0$. It can be written as
  \beq\label{cansol0}
  \Phi^{(0)}\lb z\rb=\Psi^{(0)}\lb z\rb z^{\theta_0\sigma_3},
  \eeq
  where $\Psi^{(0)}\lb z\rb$ is holomorphic and invertible on $\Cb$. The construction of canonical solutions at $z=\infty$ is more intricate. 
  There exists a \textit{formal} solution of \eqref{lisys} of the form
  \beq\label{solform}
  \Phi^{(\infty)}_{\mathrm{form}}\lb z\rb=\Psi^{(\infty)}\lb z\rb \exp\left\{\lb\tfrac{z^4}{8}+\tfrac{sz^2}{2}-\theta_\infty\ln z\rb\sigma_3\right\},
  \eeq
  where 
  \beq\label{thetainf}
  \theta_\infty=u_1v-uv_1+2s uv-u^2v^2-\theta_0, 
  \eeq
  and the coefficients of the formal series $\Psi^{(\infty)}\lb z\rb=\mathbb 1+\sum_{k=1}^{\infty}\Psi^{(\infty)}_k z^{-k}$ are uniquely determined by \eqref{lisys}. The canonical solutions $\Phi^{(\infty)}_{\ell}\lb z\rb$, $\ell\in\mathbb Z$ are uniquely specified by the asymptotic condition $\Phi^{(\infty)}_{\ell}\lb z\rb\simeq\Phi^{(\infty)}_{\mathrm{form}}\lb z\rb$ as $z\to\infty$ inside the Stokes sectors $\arg z\in\lb\frac{\lb2\ell-3\rb\pi}{8}, \frac{\lb2\ell+1\rb\pi}{8}\rb$. The canonical solutions at $0$ and $\infty$ are connected by
  \beq\label{connstokes}
  \Phi^{(\infty)}_{1}\lb z\rb=\Phi^{(0)}\lb z\rb E,\qquad
  \Phi^{(\infty)}_{\ell+1}\lb z\rb=\Phi^{(\infty)}_{\ell}\lb z\rb S_{\ell}.
  \eeq
  The monodromy data for the linear system \eqref{lisys} consist of
  (i) the exponent $\theta_0$ of local monodromy at $z=0$, 
  (ii)~the exponent $\theta_\infty$ of formal monodromy at $z=\infty$,  (iii) Stokes matrices $S_\ell$ relating canonical solutions in different sectors at $z=\infty$ and
  (iv) connection matrix $E$ relating canonical solutions at different singular points.
  
  According to the general scheme of \cite{JMU}, one may now consider $s$ in \eqref{solform} as a deformation parameter and construct a smooth family of systems \eqref{lisys} parametrized by $s$ and sharing the same monodromy. The last requirement leads to the deformation equation 
  \beq
  \partial_s\Phi=B\lb z\rb\Phi,\qquad
  B\lb z\rb=\lb\begin{array}{cc}
  \frac{z^2}{2}+uv & iz u\\ iz v &  -\frac{z^2}{2}-uv
  \end{array}\rb.
  \eeq
   The consistency condititon $\partial_zB-\partial_sA=\left[A,B\right]$ then yields a system of nonlinear ODEs for $u$, $u_1$, $v$, $v_1$ viewed as functions of $s$: 
   \begin{subequations}
   \begin{gather}
   u_1=u',\qquad v_1=v',\\
   \label{2ndeq}
   u''=2u'\lb uv-s\rb +u\lb 4suv-1-2\theta_0\rb,\\
   v''=2v'\lb s-uv\rb +v\lb 4suv+1-2\theta_0\rb. 
   \end{gather}
   \end{subequations}
   The last equation is actually redundant as it can be obtained from the previous two and \eqref{thetainf}. It then becomes straightforward to check that the function $q\lb s\rb=uv$ satisfies Painlev\'e IV equation \eqref{cP4} with parameters $\alpha=\frac{3\theta_0-\theta_\infty}{2}$, $\beta=-\frac{\lb\theta_0+\theta_\infty\rb^2}{2}$ which encode the local monodromy at $z=0$ and formal monodromy around $z=\infty$.
   
    The Stokes data are conserved quantities for Painlev\'e IV. We are now going to describe them in more detail, following \cite{ItsKapaev,Kapaev}. The Stokes matrices $S_\ell$ have the standard triangular form
  \beq
  S_{2\ell-1}=\lb \begin{array}{cc}
  1 & \eta_{2\ell-1} \\ 0 & 1
  \end{array}\rb,\qquad
   S_{2\ell}=\lb \begin{array}{cc}
    1 & 0 \\ \eta_{2\ell} & 1
    \end{array}\rb,
  \eeq
  and satisfy the quasiperiodicity relation $S_{\ell+4}=\sigma_3e^{i\pi\theta_\infty\sigma_3}S_{\ell}e^{-i\pi\theta_\infty\sigma_3}\sigma_3$, which gives $\eta_{\ell+4}=-\eta_\ell e^{2\pi i \lb -1\rb^{\ell+1}\theta_\infty}$. In combination with \eqref{cansol0}--\eqref{connstokes}, the quasiperiodicity also implies a relation between $S_\ell$'s and the connection matrix $E$: 
  \beq
  S_1S_2S_3S_4\sigma_3 e^{i\pi\theta_\infty\sigma_3}=E^{-1}\sigma_3e^{-i\pi\theta_0\sigma_3}E.
  \eeq
  Comparing the traces of both sides of this relation, one finds that
  \beq\label{ChV0}
  e^{i\pi\theta_\infty}\left[\lb 1+\eta_1\eta_2\rb \lb 1+\eta_3\eta_4\rb+\eta_1\eta_4\right]-e^{-i\pi\theta_\infty}
  \lb 1+\eta_2\eta_3\rb=-2i\sin\pi\theta_0.
  \eeq
  It follows that among the Stokes factors $\left\{\eta_\ell\right\}_{\ell\in\mathbb Z}$ at most $3$ are independent. Moreover, conjugating the fundamental solution $\Phi\lb z\rb=\Phi^{(0)}\lb z\rb$ by a constant diagonal matrix,
  $\Phi\mapsto e^{\kappa\sigma_3}\Phi e^{-\kappa\sigma_3}$, leads to the transformation
  \beq
  \eta_{2\ell-1}\mapsto e^{2\kappa}\eta_{2\ell-1}, \qquad
  \eta_{2\ell}\mapsto e^{-2\kappa}\eta_{2\ell},\qquad u\mapsto
  e^{2\kappa}u, \qquad v\mapsto
    e^{-2\kappa}v, 
  \eeq
  so that $q\lb s\rb$ remains unchanged under simultaneous rescaling of all $\eta_{2\ell-1}$ and $\eta_{2\ell}^{-1}$ by the same amount. The remaining (at most) 2 parameters encode PIV initial conditions.  
  The Stokes matrices $S_\ell$ and Stokes factors $\eta_\ell$ may be regarded as functions of $s$, $u$, $v$, $\theta_0$. There is a useful symmetry \cite[Eq. (13)]{Kapaev}
   \beq\label{slz4}
   \eta_{\ell\pm1}\lb \pm i s,e^{\pm\frac{i\pi}{4}}v,e^{\pm\frac{i\pi}{4}}u,-\theta_0\rb=
   -e^{\mp i\pi\lb -1\rb^\ell \theta_\infty/2}\eta_\ell\lb s,u,v,\theta_0\rb
   \eeq
   Given the large $|s|$ asymptotics of $q\lb s\rb$ on the ray $\arg s=\phi_0$ in terms of monodromy, the last formula allows to solve the asymptotic problem on the rays $\arg s=\phi_0+\frac{\pi\ell}{2}$, $\ell=1,2,3$.
   \begin{defin}
   The monodromy data will be called generic if $\theta_0\notin\mathbb Z+\frac12$ and the following conditions are satisfied for $\ell=0,\ldots,3$:
   \beq\label{genmon}
   1+\eta_{\ell+1}\eta_{\ell+2}\ne 0,\qquad 
   \lb 1+\eta_{\ell}\eta_{\ell+1}\rb\lb 1+\eta_{\ell+1}\eta_{\ell+2}\rb\ne 1.
   \eeq
   \end{defin}
 
  In order to connect the above to self-similar LIA solutions, one may use the function $q\lb s\rb$ given in \eqref{qdef}. The parameters of PIV can then be identified as
  \beq
  \theta_0=\tfrac12+ia,\qquad \theta_\infty=-\tfrac12+i\varepsilon.
  \eeq
  For generic monodromy, the asymptotics of $q\lb s\rb$ as $s\to\infty$ along the rays $\arg s=\frac{\pi\lb2\ell-1\rb}{4}$, $\ell=0,\ldots,3$  is described by Theorem~5 in \cite{Kapaev}. Comparing the result with e.g. the  estimate \eqref{asq}, we find that
    \beq\label{AsPar}
    \begin{gathered}
    e^{-2\pi\lb \omega_+-\frac{\varepsilon}{3}\rb}=1+\eta_1\eta_2,\qquad e^{-2\pi\lb \omega_--\frac{\varepsilon}{3}\rb}=1+\eta_3\eta_4,\\
    1-e^{i\rho_+}=\lb1+\eta_0\eta_1\rb
    \lb1+\eta_1\eta_2\rb,\qquad
   1-e^{i\rho_-}=\lb 1+\eta_2\eta_3\rb\lb 1+\eta_3\eta_4\rb. 
    \end{gathered}
    \eeq
   The connection formulas  \eqref{connfI}--\eqref{connfII} in Theorem~\ref{theob} can now be deduced from the quasiperiodicity relation $\eta_0=\eta_4e^{-2\pi\varepsilon}$ and the monodromy equation \eqref{ChV0} rewritten as
    \beq\label{ChV}
     e^{-\pi\varepsilon}\left[\lb1+\eta_1\eta_2\rb \lb1+\eta_3\eta_4\rb+\eta_1\eta_4\right]+e^{\pi\varepsilon}\lb1+\eta_2\eta_3\rb=2\cosh \pi a.
    \eeq
    These formulas only involve the asymptotics of $q\lb s\rb$ as $s\to\pm e^{-\frac{i\pi}{4}}\infty$ and remain valid even if the conditions \eqref{genmon} are relaxed for $\ell=1,3$.

  \subsection{Odd and mixed symmetry solutions\label{subsecodd}}
  We are now going to discuss in more detail the asymptotic and connection properties of the curvature and torsion for two classes of symmetric self-similar solutions of \eqref{SSs}, dubbed odd and mixed cases in \cite{GV}. Choose an orthonormal frame $\lb\mathbf e_1,\mathbf e_2,\mathbf e_3\rb$ such that $\ba=\lb0,0,a\rb$. In this frame, symmetric solutions satisfy 
  \begin{align}
  \bG\lb -s\rb=\begin{cases}
  -\bG\lb s\rb,\qquad\qquad &\text{(odd)}\\
  \operatorname{diag}\lb1,1,-1\rb\bG\lb s\rb.\qquad\qquad &\text{(mixed)}
  \end{cases}
  \end{align}
  Since the functions $\sigma\lb s\rb=aG_3\lb s\rb$, $\sigma''\lb s\rb$ are odd in both cases,  we have $\sigma\lb0\rb=\sigma''\lb 0\rb=0$. Substituting these initial conditions into $\sigma$-PIV equation \eqref{sP4}, it follows that $\sigma'\lb 0\rb$ is equal to $\varepsilon$ or $\pm a$. The first option corresponds to the odd case and the latter two to the mixed one, cf \cite[Section~5]{GV}. We denote the corresponding solutions by $\sigma_\alpha\lb s\rb$, with $\alpha\in\left\{\varepsilon,a,-a\right\}$.
  
  The parameter $\varepsilon$ is constrained by the conditions $|\sigma'\lb s\rb|\leq a$, $\varepsilon-\sigma'\lb s\rb\ge 0$ coming from $\|\bG'\|=1$, ${\varepsilon=\|\bG''\|^2+\sigma'}$. Therefore, symmetric solutions $\sigma_\varepsilon\lb s\rb$, $\sigma_{\pm a}\lb s\rb$ can exist only for $|\varepsilon|\leq a$, $\varepsilon \ge\pm a$, respectively. We would like to describe the asymptotics of these solutions as $s\to\pm\infty$, as this happens to be one of the rare cases where such behavior can be explicitly related to the Cauchy type initial conditions at a finite point (here $s=0$).
  
  The crucial point is that the parity properties allow to express the monodromy data in terms of the parameters $\varepsilon,a$ of Painlevé IV. For example, from the asymptotics \eqref{sigmaas} as $s\to\pm\infty$ it follows that 
  \beq
  \omega_+=\omega_-,\qquad \rho_+=\rho_-,
  \eeq
  which translates into conditions on the Stokes coefficients,
  \beq
  \eta_1\eta_2=\eta_3\eta_4,\qquad \eta_1\eta_4=e^{2\pi \varepsilon}\eta_2\eta_3.
  \eeq
  The equation \eqref{ChV} then produces four solutions for $\omega=\omega_\pm$:
  \beq
  X\equiv e^{ \frac{2\pi\lb\varepsilon-3\omega\rb}{3}}\in\left\{-e^{\pi\varepsilon}\pm 2e^{\frac{\pi\varepsilon}{2}}\cosh\tfrac{\pi a}{2},
  e^{\pi\varepsilon}\pm 2e^{\frac{\pi\varepsilon}{2}}\sinh\tfrac{\pi a}{2}\right\}.
  \eeq
  To each of them is also assigned a unique solution for $e^{i\rho_\pm}$ but the corresponding expressions are omitted for simplicity.
  Not all of these solutions are actually relevant, in particular, because of the condition $\varepsilon\ge 3\omega$:
  \begin{itemize}
  \item Solution  $X_1=-e^{\pi\varepsilon}- 2e^{\frac{\pi\varepsilon}{2}}\cosh\tfrac{\pi a}{2}$ is negative and therefore may be discarded.
  \item Solution $X_2=-e^{\pi\varepsilon}+ 2e^{\frac{\pi\varepsilon}{2}}\cosh\tfrac{\pi a}{2}$ satisfies $X_2\ge1$ only for $|\varepsilon|\leq a $.
  \item Solution $X_3=e^{\pi\varepsilon}- 2e^{\frac{\pi\varepsilon}{2}}\sinh\tfrac{\pi a}{2}$ satisfies $X_3\ge 1$ only for $\varepsilon\ge a$.
  \item Solution $X_4= e^{\pi\varepsilon}+ 2e^{\frac{\pi\varepsilon}{2}}\sinh\tfrac{\pi a}{2}$ always satisfies $X_4\ge 1$.
  \end{itemize}
  
  Thus we have at most two admissible solutions ($X_2$ and $X_4$) for $|\varepsilon|<a$ as well as at most two admissible solutions ($X_3$ and $X_4$) for $\varepsilon>a$. On the other hand, we know that symmetric solutions $\sigma_\varepsilon\lb s\rb$ and $\sigma_{-a}\lb s\rb$ do exist for $|\varepsilon|<a$, while $\sigma_a\lb s\rb$ and $\sigma_{-a}\lb s\rb$ do exist for $\varepsilon>a$. One may therefore expect the identification
  \beq\label{matching}
  X_2\mapsto\sigma_{\varepsilon}\lb s\rb,\qquad X_3\mapsto\sigma_{a}\lb s\rb,\qquad X_4\mapsto\sigma_{-a}\lb s\rb.
  \eeq
  This matching is quite plausible: indeed, there is no reason to expect any discontinuous behavior of the mixed solution $\sigma_{-a}\lb s\rb$ in $\varepsilon$ when the latter crosses the value $\varepsilon=a$ (which is why $\sigma_{-a}\lb s\rb$ should correspond to $X_4$). Moreover, the matching \eqref{matching} is readily confirmed by numerics.  However, since we were unable to  justify it rigorously, the following result is formulated as a conjecture rather than a theorem. 
  
  \begin{conj}
  The asymptotics of the curvature and torsion for the symmetric solutions $\sigma_{\pm a,\varepsilon}\lb s\rb$ is described by the formulas of Theorem~\ref{theob} with parameters
  \begin{enumerate}
  \item Odd solution, $\mathsf C\lb 0\rb=0$, $|\varepsilon|<a$:
  \begin{subequations}
  \beq\label{oddmon}
  \omega_\pm=\tfrac{\varepsilon}{12}-\tfrac1{2\pi}\ln\lb
  2\cosh\tfrac{\pi a}{2}-e^{\frac{\pi\varepsilon}{2}}\rb,\qquad \Re\rho_\pm=\pi.
  \eeq
  \item Mixed solution, $\mathsf C^2\lb 0\rb=\varepsilon-a>0$:
    \beq
    \omega_\pm=\tfrac{\varepsilon}{12}-\tfrac1{2\pi}\ln\lb
    e^{\frac{\pi\varepsilon}{2}}-
          2\sinh\tfrac{\pi a}{2}\rb,\qquad \Re\rho_\pm=\pi.
    \eeq
  \item  Mixed solution, $\mathsf C^2\lb 0\rb=\varepsilon+a>0$:
      \beq
      \omega_\pm=\tfrac{\varepsilon}{12}-\tfrac1{2\pi}\ln\lb e^{\frac{\pi\varepsilon}{2}}+
      2\sinh\tfrac{\pi a}{2}\rb,\qquad \Re\rho_\pm=0.
      \eeq  
  \end{subequations}
  \end{enumerate}
  In particular, the limiting values of the curvature scaling function are $\mathsf C^2\lb+\infty\rb=\mathsf C^2\lb-\infty\rb=\frac{2\lb\varepsilon-3\omega_\pm\rb}{3}$.
  \end{conj}
  
  \begin{rmk} A particular odd solution singled out in \cite{GV} describes an asymptotically planar spiral filament. It corresponds to $G_3\lb\pm\infty\rb=0$, which implies (cf \eqref{sigmaas0}) that $\varepsilon+6\omega_\pm=0$. In combination with the 1st equation in \eqref{oddmon}, this in turn gives $\varepsilon=\frac2\pi\ln\cosh\frac{\pi a}{2}$, thereby fixing  the relevant initial conditions at $s=0$ in terms of the only remaining parameter $a$. Namely, the parameter $\delta=G_3'\lb 0\rb$ in \cite[eq. (164)]{GV} is equal to $\delta=\frac2{\pi a}\ln\cosh\frac{\pi a}{2}$. One may compare e.g. the value $\delta\approx 0.95587$ found for $a=10$ with the one used to produce Fig. 2 in \cite{GV}. 
  \end{rmk}

 \section{Asymptotics and explicit solution for $a=0$\label{sec_ae0}}
  In principle, it should be possible to  prove Theorem~\ref{theoc} by direct differentiation of the representations \eqref{1F1eqs} and application of various hypergeometric function identities. Below we outline a more intuitive bottom-up approach.
  
  \begin{prop} The solution of \eqref{ae0eq} satisfying the initial conditions \eqref{ae0ic} is given by
  \begin{subequations}
  \beq\label{Gpe}
  G'_j\lb s\rb = 1-\kappa\lb\lambda_{+,j},\lambda_{-,j}\rb^{-1}\prod_{\nu=\pm}\lb\sum_{\mu=\pm}
    e^{\mu\lambda_{\nu,j}}D_{-\frac{i\nu\varepsilon}{2}}\lb\frac{\mu e^{i\pi\nu/4}s}{\sqrt2}\rb\rb ,\qquad j=1,2,3
  \eeq
  where $D_\alpha\lb z\rb$ is the parabolic cylinder function and
  \begin{gather}
   \label{kappae}
  \kappa\lb \lambda_+,\lambda_-\rb=e^{\frac{\pi \varepsilon}{4}}\cosh\lb\lambda_{+}+\lambda_{-} \rb
  +e^{-\frac{\pi \varepsilon}{4}}\cosh\lb\lambda_{+}-\lambda_{-} \rb,\\
  \label{lambdae}
  \lambda_{\pm,1}=\frac{i\pi}2,\qquad
  \tanh\lambda_{\pm,2}=\pm i\tanh\lambda_{\pm,3}=-\frac{2e^{\pm\frac{i\pi}{4}}}{\sqrt\varepsilon}\frac{\Gamma\lb1\pm\frac{i\varepsilon}{4}\rb}{
    \Gamma\lb\frac12\pm\frac{i\varepsilon}{4}\rb}.
  \end{gather}
  \end{subequations}
  \end{prop}

  \pf The first step of the proof mimics the derivation of PIV in the $a\ne 0$ case. Fix a constant unit vector $\mathbf{e}$ and denote $\zeta=\mathbf e\cdot \bG$. We have
  \beq
  \|\mathbf{e}\|^2=1,\quad \|\bG\|^2=s^2+4\varepsilon,\quad 
  \|\bG'\|^2=1, \qquad \mathbf e\cdot \bG=\zeta,\qquad 
  \mathbf e\cdot \bG'=\zeta',\qquad \bG\cdot\bG'=s.
  \eeq
  Also,  $\bG''=\frac12\bG\wedge\bG'$ and therefore $\mathbf e\cdot\lb \bG\wedge\bG'\rb=2\zeta''$. Then, setting $\mathbf a=\mathbf e$, $\mathbf b=\bG$, $\mathbf c=\bG'$ in the identity
  \eqref{determ}, it can be easily  checked that $\zeta$ satisfies the equation
   \beq\label{sP4lim}
     \lb \zeta''\rb^2+ \tfrac14\lb s\zeta'-\zeta\rb^2=\varepsilon\lb1-\zeta'^2\rb.
    \eeq 
  The same result can also be formally obtained from the $\sigma$-Painlev\'{e} IV equation \eqref{sP4} by setting $\sigma=a\, \zeta$ and sending $a$ to $0$. It should be emphasized, however, that $\mathbf e$ can be chosen arbitrarily in the above.  
  
  In order to integrate \eqref{sP4lim},  introduce the combinations
  \beq\label{qpm}
  q_\pm=\frac{\zeta''\pm\frac i2\lb s\zeta'-\zeta\rb}{1-\zeta'}.  
  \eeq
  This substitution is inspired by the formulas \eqref{qpdef}  relating the standard and $\sigma$-form of Painlev\'{e} IV.  It is straightforward to verify that $q_\pm$ satisfy Riccati equations
  \beq
  2q_\pm'=q_\pm^2\pm i s q_\pm+\varepsilon.
  \eeq
  The substitution $q_{\pm}=-2\frac{d}{ds}\ln f_{\pm}$ reduces them to 2nd order linear ODEs, $4f_\pm''\mp 2i sf_\pm'+\varepsilon f_\pm=0$, which are equivalent to Hermite differential equation.
  The respective general solutions for $q_\pm$ may then be written as
  \beq
  q_{\pm}\lb s\rb=-2\frac{d}{ds}\ln\sum_{\mu=\pm} e^{\pm\frac{is^2}{8}+\mu\lambda_\pm}D_{\mp\frac{i\varepsilon}{2}}\lb\frac{\mu e^{\pm i\pi/4}s}{\sqrt2}\rb,
  \eeq
 where $\lambda_\pm$ are integration constants. Now, add equations \eqref{qpm} for $q_\pm$ and integrate the result:
  \beq\label{DD}
  \zeta'\lb s\rb=1-\kappa^{-1}\prod_{\nu=\pm}\lb\sum_{\mu=\pm}
  e^{\mu\lambda_\nu}D_{-\frac{i\nu\varepsilon}{2}}\lb\frac{\mu e^{i\pi\nu/4}s}{\sqrt2}\rb\rb.
  \eeq
  We have thus shown that the solution of \eqref{sP4lim} necessarily has the form \eqref{DD}. However, the integration constants $\lambda_\pm$, $\kappa$ are not independent. The relation between them can be found by rewriting \eqref{sP4lim} in the form
  $q_+ q_-=\varepsilon\displaystyle\frac{1+\zeta'}{1-\zeta'}$ and evaluating both sides at $s=0$, which implies that $\kappa=\kappa\lb\lambda_+,\lambda_-\rb$ is given by \eqref{kappae}.

  Setting $\mathbf e=\mathbf e_j$ ($j=1,2,3$), we see that each of $G'_j\lb s\rb$ is given by \eqref{DD}. It remains to choose the integration constants $\lambda_{\pm,j}$ as to satisfy the initial conditions \eqref{ae0ic}. The details of this cumbersome but straightforward computation are omitted; we only note that e.g. $\zeta'\lb 0\rb=0$ implies that     $\tanh \lambda_+\tanh\lambda_-=\coth\frac{\pi \varepsilon}{4}$.  The values of $\tanh\lambda_{\pm,j}$ are uniquely fixed by  $G'_j\lb 0\rb$, $G''_j\lb 0\rb$ for $j=1,2$;  for $j=3$ there remains a two-fold ambiguity which can be lifted by requiring consistency of the full system \eqref{ae0eq}. The solution \eqref{DD} depends on the initial conditions only via exponentials $e^{2\lambda_{\pm}}$ (equivalently, $\tanh\lambda_{\pm}$), therefore the choice of solution of \eqref{lambdae} for $\lambda_{\pm,j}$ is not essential. 
    \epf

  In the derivation leading to \eqref{sP4lim}, one could use instead of $\mathbf  e$ a complex vector of zero norm, e.g. $\tilde{\mathbf e}=\mathbf e_2+i\mathbf e_3$. The complex-valued function $\tilde\zeta\lb s\rb=\tilde{\mathbf e}\cdot \bG\lb s\rb$ would then satisfy the equation
     \beq\label{sP4lim2}
       \lb \tilde\zeta''\rb^2+ \tfrac14\lb s\tilde\zeta'-\tilde\zeta\rb^2+\varepsilon\tilde\zeta'^2=0.
      \eeq 
  A computation similar to the one carried in the previous proof yields the general solution of \eqref{sP4lim2}:
  \beq\label{G23}
  \tilde{\zeta}'\lb s\rb=\tilde\kappa^{-1}\prod_{\nu=\pm}\lb\sum_{\mu=\pm}
    e^{\mu\tilde\lambda_\nu}D_{-\frac{i\nu\varepsilon}{2}}\lb\frac{\mu e^{i\pi\nu/4}s}{\sqrt2}\rb\rb,
  \eeq
  where, in general, the integration constants $\tilde\kappa$, $\tilde\lambda_\pm$ satisfy $\tanh\tilde\lambda_+\tanh\tilde\lambda_-=-\coth\frac{\pi\varepsilon}{4}$. For the initial conditions we are interested in, i.e. $\tilde\zeta'\lb 0\rb=0$,  we have  $\tilde\lambda_+=\frac{i\pi}{2}$, $\tilde\lambda_-=0$. The same remains true if we start from $\tilde{\mathbf e}=\mathbf e_2-i\mathbf e_3$, except that the values of $\tilde \lambda_+$ and $\tilde\lambda_-$ have to be swapped.
  
  Observe that the formula \eqref{Gpe} for $G_1'\lb s\rb$ as well as representations \eqref{G23} for $G_2'\lb s\rb\pm i G_3'\lb s\rb$ involve only sums and differences $D_\alpha\lb z\rb\pm D_\alpha\lb -z\rb$ of the parabolic cylinder functions. These combinations can be rewritten in terms of the confluent hypergeometric functions using that
  \begin{subequations}
  \begin{align}
  D_\alpha\lb z\rb+D_\alpha\lb -z\rb=&\,\frac{2^{1+\frac\alpha 2}\sqrt{\pi}}{\Gamma\lb\frac{1-\alpha}{2}\rb}\,e^{-\frac{z^2}{4}}{\,}_1F_1\lb-\tfrac{\alpha}{2},\tfrac12,\tfrac{z^2}{2}\rb,\\
  D_\alpha\lb z\rb-D_\alpha\lb -z\rb=&\,-\frac{2^{\frac{3+\alpha} 2}\sqrt{\pi}}{\Gamma\lb-\frac{\alpha}{2}\rb}\,z\, e^{-\frac{z^2}{4}}{\,}_1F_1\lb\tfrac12-\tfrac{\alpha}{2},\tfrac32,\tfrac{z^2}{2}\rb,
  \end{align}
  \end{subequations}
  which ultimately yields compact formulas \eqref{1F1eqs} of Theorem~\ref{theoc}.  
    
     The behavior of $\bG'\lb s\rb$ as $s\to\pm\infty$ can now be easily characterized using the explicit hypergeometric expressions and the known asymptotics of $_1F_1\lb\alpha,\gamma,z\rb$ as $z\to\pm i\infty$. In particular, one finds the following result (cf \cite[Theorem~1]{GRV}).
  \begin{prop} As $s\to+\infty$, we have
  \begin{subequations}
  \begin{align}
  G'_1\lb s\rb=&\,e^{-\frac{\pi\varepsilon}{2}}+
      \frac{2\sqrt{\varepsilon\lb 1-e^{-\pi\varepsilon}\rb}}{s}\,
      \cos\lb \Omega\lb s\rb-\beta_1-\beta_2\rb+O\lb s^{-2}\rb,\\
  G'_2\lb s\rb\pm i   G'_3\lb s\rb=&\,   
  \sqrt{ 1-e^{-\pi\varepsilon}}\,e^{\pm i\lb\beta_1-\beta_2\rb} 
  +\frac{\sqrt{\varepsilon}}{s}\left[\lb 1-e^{-\frac{\pi\varepsilon}{2}}\rb e^{\mp i\Omega\lb s\rb\pm2i\beta_1}-\lb 1+e^{-\frac{\pi\varepsilon}{2}}\rb e^{\pm i\Omega\lb s\rb\mp 2i\beta_2}\right]+O\lb s^{-2}\rb,
  \end{align}
      where 
      \beq
      \Omega\lb s\rb=\tfrac{s^2}{4}+\varepsilon\ln\tfrac{s}{2},\qquad
      \beta_1=\arg \Gamma\lb1+\tfrac{i\varepsilon}{4}\rb,\qquad
      \beta_2=\arg \Gamma\lb\tfrac12+\tfrac{i\varepsilon}{4}\rb-\tfrac{\pi}4.
      \eeq 
 \end{subequations}      
   \end{prop} 
   
    The asymptotics of $\bG'\lb s\rb$ as $s\to -\infty$ follows from the parity properties of different components. The asymptotic directions at $\pm \infty$ are given by 
    \beq\mathbf T_{\pm}=\lb e^{-\frac{\pi\varepsilon}{2}},\pm\sqrt{ 1-e^{-\pi\varepsilon}}\cos\lb\beta_1-\beta_2\rb,\pm\sqrt{ 1-e^{-\pi\varepsilon}}\sin\lb\beta_1-\beta_2\rb\rb.
    \eeq
    The angle between them monotonously increases from $0$ to $\pi$ with the growth of curvature parameter $\varepsilon$, which is clearly seen from $\mathbf T_{+}\cdot \mathbf T_{-}=2e^{-\pi\varepsilon}-1$.

    \section{Discussion}

 In this work, we have studied self-similar solutions of the vortex filament evolution, which can also be formulated as the Landau-Lifshitz equation for an isotropic continuous spin chain --- a one-dimensional ferromagnet. The self-similar dynamics is governed by Painleve IV equation with two real parameters. 
 We explicitly describe the  asymptotics of solutions as well as the corresponding connection formulae.

 The Cauchy problem for the original equations is formulated by fixing profiles at the initial moment of time.  
 The case $a=0$ corresponds to a corner-like initial profile of the filament or to the domain-wall spin-configurations in the LL ferromagnet. 
 It was recently checked that numerical solutions with the smoothed domain-wall profile approach the self-similar solutions after a certain amount of time \cite{GMI}. 
 It is interesting to note, however, that self-similar solutions cannot fully capture anti-parallel domain-wall profiles $T_3\lb x\to\pm \infty\rb = \pm 1$.
 Indeed, if it were the case then at large times the half-space magnetization would grow as a square root of time 
 $M\lb t\rb= \int_0^\infty \lb 1-T_3\lb x\rb\rb dx\simeq \int_0^\infty \lb 1-T_3\lb x/\sqrt{t}\rb\rb dx \sim \sqrt{t}$. It was observed, however, that the growth gets a logarithmic enhancement $M(t)\sim \sqrt{t}\,\ln t$ \cite{GMI}. 
 This situation is reminiscent of the asymptotic behavior of classical integrable systems whose scattering data contain real poles of the transition amplitude \cite{AS}.

  In general, self-similar solutions describe certain regions of the long-time asymptotic behavior of integrable non-linear wave equations with arbitrary initial profiles (see, for instance, \cite{Deift}). The relevant ordinary differential equation often belong to Painlev\'{e} class. Nevertheless, the complete significance of the self-similar solutions investigated in the present manuscript is yet to be clarified.\vspace{0.2cm}
   
    \noindent
    { \small \textbf{Acknowledgements}.  O.~G. acknowledges the support from the European Research Council under ERC Advanced grant 743032 DYNAMINT.}


\begin{thebibliography}{100}
 	
 
 \bibitem{AH}
 R. J. Arms, F. R. Hama, {\it Localized-induction concept on a curved vortex and motion of an elliptic vortex ring},
 Phys. Fluids \textbf{8}, (1965), 553--559.
 \bibitem{AS}
 M. J. Ablowitz,  H. Segur, {\it Asymptotic Solutions of the Korteweg-de Vries Equation}, Stud. Appl. Math. \textbf{57}, (1977), 13--44.
 \bibitem{BV}
 V. Banica, L. Vega, {\it The initial value problem for the binormal flow with rough data}, Ann. Sci. ENS \textbf{48}, (2015), 1423--1455;
 arXiv:1304.0996 [math.AP].
 \bibitem{BCHM}
 A. P. Bassom, P. A. Clarkson, A. C. Hicks, J. B. McLeod, \textit{Integral solutions and exact solutions for the fourth Painlev\'{e} equation}, Proc. Roy. Soc. London \textbf{A437}, (1992), 1--24. 
 \bibitem{Bet}
 R. Betchov, {\it On the curvature and torsion of an isolated vortex filament}, J. Fluid Mech. \textbf{22}, (1965), 471--479. 
 \bibitem{BP}
 M. Boiti, F. Pempinelli, {\it Nonlinear Schr\"odinger equation, B\"acklund transformations and Painlev\'{e} transcendents}, Nuovo Cim. \textbf{59B}, (1980), 40--58.
 \bibitem{BLMST} G. Bonelli, O. Lisovyy, K. Maruyoshi,
  A. Sciarappa, A. Tanzini, {\it On Painlev\'{e}/gauge theory correspondence}, Lett. Math. Phys. \textbf{107}, (2017), 2359--2413; arXiv:1612.06235 [hep-th]. 
 \bibitem{Can}
 M. Can, {\it On the relations between nonlinear Schr\"odinger equation and Painlev\'{e} IV equation}, Nuovo Cim. \textbf{106B}, (1991), 205--207.
 \bibitem{Clarkson}
 P. Clarkson, {\it The fourth Painlev\'{e} equation}, in ``Differential Algebra and Related Topics'' (eds. L. Guo, W. Y. Sit), World Scientific, Singapore, (2008); https://kar.kent.ac.uk/23090/
 \bibitem{DaRios} L. S. Da Rios, {\it On the motion of an unbounded fluid with a vortex filament of any shape}, Rend. Circ. Mat. Palermo \textbf{22}, (1906), 117--135. 
 \bibitem{Deift}P. A. Deift, A. R. Its, X. Zhou, \textit{Long-time asymptotics for integrable nonlinear wave equations}, in ``Important Developments in Soliton Theory'' (eds. A. S. Fokas, V. E. Zakharov), Springer, Berlin, Heidelberg,
 (1993), pp. 181--204.
 \bibitem{Dunne}
 G. V. Dunne, \textit{Resurgence, Painlev\'{e} equations and conformal blocks}, arXiv:1901.02076v1 [hep-th].  
 \bibitem{FIKN} A. S. Fokas, A. R. Its, A. A. Kapaev, V. Yu. Novokshenov, \textit{Painlev\'e transcendents:
  the Riemann-Hilbert approach}, Mathematical Surveys and Monographs~\textbf{128}, AMS, Providence,
  RI, (2006).
  \bibitem{GIL12}
   O. Gamayun, N. Iorgov, O. Lisovyy,  \textit{Conformal field theory of Painlev\'e~VI},
   J. High Energ. Phys. (2012) 2012: 38; arXiv:1207.0787 [hep-th].
   \bibitem{GIL13}
    O. Gamayun, N. Iorgov, O. Lisovyy,  \textit{How instanton combinatorics solves Painlev\'{e}~VI, V and III's},
     J.~Phys.~\textbf{A46}, (2013), 335203;
      {arXiv:1302.1832 [hep-th]}.     
      \bibitem{GMI}  O. Gamayun, Y. Miao, E. Ilievski,  \textit{Domain wall dynamics in the Landau--Lifshitz magnet and the classical-quantum correspondence of spin transport}, {arXiv:1901.08944  [cond-mat]}.
 \bibitem{GRV}
 S. Guti\'errez, J. Rivas, L. Vega, \textit{Formation of singularities and self-similar vortex motion under the localized induction approximation}, Comm. Part. Diff. Eqs. \textbf{28}, no. 5--6, (2003), 927--968.
  \bibitem{GV}
  S. Guti\'errez, L. Vega, \textit{Self-similar solutions of the localized induction approximation: singularity formation}, Nonlinearity~\textbf{17}, (2004), 2091--2136; arXiv:math/0404291 [math.AP].
   \bibitem{GV2}
   S. Guti\'errez, L. Vega, \textit{On the stability of self-similar solutions of 1D cubic Schr\"odinger equations}, Math. Ann.~\textbf{356}, (2013), 259--300; arXiv:1103.5403 [math.AP].
  \bibitem{Hasimoto}
  H. Hasimoto, {\it A soliton on a vortex filament}, J. Fluid Mech.~\textbf{51}, (1972), 477--485.
 \bibitem{ItsKapaev}
  A. R. Its, A. A. Kapaev, \textit{Connection formulae for the fourth Painlev\'e transcendent: Clarkson-McLeod solution}, J. Phys. \textbf{A31}, (1998), 4073--4113.
 \bibitem{ILT}
 A. Its, O. Lisovyy, Yu. Tykhyy, \textit{Connection problem for the sine-Gordon/Painlev\'{e} III tau function and irregular conformal blocks},
 Int. Math. Res. Not. \textbf{2015}, Issue 18, (2015), 8903--8924;  arXiv:1403.1235 [math-ph].
 \bibitem{JMU}
 M. Jimbo, T. Miwa, K. Ueno, {\it Monodromy preserving deformation of linear ordinary differential equations with rational coefficients I. General theory and $\tau$-function}, Physica~\textbf{2D}, (1981), 306--352. 
 \bibitem{JMII}
 M. Jimbo, T. Miwa, {\it Monodromy preserving deformation of linear ordinary differential equations with rational coefficients. II}, Physica~\textbf{2D}, (1981), 407--448.
 \bibitem{Kapaev}
 A. A. Kapaev, \textit{Global asymptotics of the fourth Painlev\'{e} transcendent}, Steklov Math. Inst. and IUPUI Preprint 6/1996, (1996);
 ftp://ftp.pdmi.ras.ru/pub/publicat/preprint/1996/06-96.ps.gz.
 \bibitem{KapaevII}
  A. A. Kapaev, \textit{Connection formulae for degenerated asymptotic solutions of the fourth Painlev\'{e} equation}, arXiv:solv-int/9805011. 
 \bibitem{Kitaev}
 A. V. Kitaev, \textit{Asymptotic description of the fourth Painlev\'{e} equation solutions on the Stokes rays analogies}, Zap. Nauchn. Sem. LOMI~\textbf{169}, (1988), 84--90.
 \bibitem{LRT}
 M. Lakshmanan, Th. W. Ruijrok, C. J. Thompson,
 \textit{On the dynamics of a continuum spin system}, Physica \textbf{A84}, (1976), 577--590.
 \bibitem{Lip1}
 T. Lipniacki, {\it Shape-preserving solutions for quantum vortex motion under localized induction approximation}, Physics of Fluids \textbf{15}, no. 6, (2003), 1381--1395.
 \bibitem{Lip2}
 T. Lipniacki, {\it Quasi-static solutions for quantum vortex motion
 under the localized induction approximation}, J. Fluid Mech. \textbf{477}, (2003), 321--337.
 \bibitem{Nagoya1}
 H. Nagoya, \textit{Irregular conformal blocks, with an application to the fifth and fourth Painlev\'{e} equations}, J.~Math. Phys.~\textbf{56}, (2015), 123505; arXiv:1505.02398 [math-ph].
 \bibitem{Nagoya2}
 H. Nagoya, \textit{Conformal blocks and Painlev\'{e} functions},  	arXiv:1611.08971 [math-ph]. 
 \bibitem{QC} 
 G. R. W. Quispel, H. W. Capel, \textit{Equation of motion for the Heisenberg spin chain}, Phys. Letts. \textbf{85A}, no. 4, (1981), 248--250.
 \bibitem{Ricca}
 R. L. Ricca, \textit{Rediscovery of Da Rios equations}, Nature \textbf{352}, (1991), 561--562.
 \end{thebibliography}
 \end{document}